\documentclass[12pt,preprint]{aastex} % one-column preprint format

\usepackage{graphicx}
\usepackage[twocolumn]{emulateapj5}

\newcommand{\halpha}{H$\alpha$}
\newcommand{\hbeta}{H$\beta$}

\shortauthors{SCHMIDT ET AL.}
\shorttitle{NEW LOW-$\dot M$ MAGNETIC BINARIES}
\slugcomment{}

\received{}
\revised{}

\begin{document}

\tolerance 10000

\title{TWO ADDITIONS TO THE NEW CLASS OF LOW ACCRETION-RATE MAGNETIC BINARIES}

\author{
Gary D. Schmidt\altaffilmark{1},
Paula Szkody\altaffilmark{2},
Arne Henden\altaffilmark{3},
Scott F. Anderson\altaffilmark{2},
Don Q. Lamb\altaffilmark{4},
Bruce Margon\altaffilmark{5},
\and
Donald P. Schneider\altaffilmark{6}
}
\vskip 10pt

\altaffiltext{1}{Steward Observatory, The University of Arizona, Tucson, AZ
85721.} \email{gschmidt@as.arizona.edu}
\altaffiltext{2}{Department of Astronomy, University of Washington, Box 351580,
Seattle, WA 98195-1580.}
\altaffiltext{3}{American Association of Variable Star Observers, 25 Birch St.
Cambridge, MA 02138.}
\altaffiltext{4}{Department of Astronomy \& Astrophysics, University of Chicago, 5640 S. Ellis Ave., Chicago, IL 60637.}
\altaffiltext{5}{Space Telescope Science Institute, 3700 San Martin Drive,
Baltimore, MD 21218.}
\altaffiltext{6}{Pennsylvania State University, Department of Physics \& Astronomy,
525 Davey Laboratory, University Park, PA 16802.}

\begin{abstract}

Two new magnetic white dwarf accretion binaries with extremely low
mass-transfer rates have been discovered in the course of the Sloan Digital Sky
Survey.  Measured magnetic fields are 42~MG and 57~MG, and one system orbits
with a period of just 82 min.  The new systems therefore significantly expand
the range in properties exhibited by the small class.  The measured accretion
rates are very low, $0.6-5\times10^{-13}~M_\sun$~yr$^{-1}$, and multiple visits
spanning more than a year confirm that this is not a short-lived
characteristic.  It is becoming increasingly clear that the low-$\dot M$
magnetic white dwarf binaries accrete by nearly complete magnetic capture of
the stellar wind from the secondary star rather than by Roche lobe overflow.
The accretion rates therefore provide some of the first realistic estimates of
the total wind loss rates from M dwarfs.  Although one or more of the eight
systems known to date may be interrupted or possibly even extinct Polars,
several lines of evidence suggest that most are pre-Polars whose evolution has
not yet brought the secondaries into contact with their Roche surfaces.
Considering the difficulties of identifying binaries over a wide range in field
strength and accretion rate, it is quite possible that the space density of
wind-accreting magnetic binaries exceeds that of the classical X-ray emitting,
Roche-lobe overflow Polars.

\end{abstract}


\section{Introduction}

Progress toward an understanding of the common envelope (CE) and post-common
envelope stages of binary star evolution has been slow, in part due to the
difficulties of identifying and characterizing populations of white dwarf + M/L
dwarf close binary systems that are sufficiently numerous and cover wide ranges
in age and mass ratio.  The particular questions and problems posed by the
(lack of) detached binaries with a magnetic degenerate component have recently
been discussed by Liebert et al. (2005). Among the cataclysmic variables (CVs)
- the eruptive phase that ensues for sufficiently close pairs - the strongly
magnetic, circularly polarized systems (AM Herculis binaries, or Polars; see,
e.g., Wickramasinghe \& Ferrario 2000) have often been more useful than the
nonmagnetic examples for studying the stellar components. In these systems the
magnetic field disrupts the formation of an accretion disk and most of the
accretion energy emerges as X-rays/EUV radiation.  Moreover, when these
binaries lapse into states of low accretion, the stellar continua are almost
uncontaminated by cyclotron emission.

Recent optical spectroscopic surveys have uncovered a population of magnetic
binaries that display remarkably isolated cyclotron harmonics and accretion
rates $<$1\% of the values typically encountered among Roche-lobe overflow CVs
(Reimers et al. 1999; Reimers \& Hagen 2000; Schwope et al. 2002; Szkody et
al.  2003; Schmidt et al. 2005a, hereafter S05).  A variety of clues, including
secondary stars that underfill their Roche lobes and unusually cool primary
star temperatures, indicates that accretion onto the white dwarfs in these
systems occurs through efficient (magnetic) capture of the stellar wind from
the low-mass secondary (Webbink \& Wickramasinghe 2005; S05).  This process was
originally explored in relation to the orbital period evolution of Polars
(Webbink \& Wickramasinghe 2002; Li et al. 1994, 1995), since angular momentum
loss via the stellar wind of the companion is generally thought to be
important for orbital periods $P\gtrsim 3$ hr.  Because the new low-accretion
rate binaries appear to have not yet entered into Roche lobe contact, they have
been interpreted as pre-magnetic CVs, or pre-Polars.  If this is correct, they
offer fresh insight into the products of post-CE evolution as well as to the
various routes followed in the formation of magnetic CVs.  In this paper we
report the discovery of two new low-accretion rate magnetic examples that
significantly broaden the range of characteristics exhibited by the class.

\section{Observations}

Like most of the prior discoveries, the two newest low-accretion rate magnetic
systems were found in the course of the Sloan Digital Sky Survey (SDSS; York et
al. 2000).  Both sources were selected for dual-channel (blue/red) fiber
spectroscopy as candidate QSOs by the automatic targeting algorithms because
their {\it ugriz\/} photometric colors (Fukugita et al. 1996) place them well
off the stellar locus (Richards et al. 2002). However, as seen in Figure 1, the
optical spectra are quite different from QSOs, with SDSS~J105905.07+272755.5
characterized by the spectrum of a late-type star and SDSS~J103100.55+202832.2
displaying a blue underlying continuum\footnote{Hereafter, objects will be
designated by SDSS~J{\em hhmm}$\pm${\em ddmm}.}.  Cyclotron harmonics and
magnetic field strengths are identified in the figure, computed under the
assumption of low plasma temperatures ($kT_e\sim1$~keV; e.g., Szkody et al.
2004).  Details of the SDSS photometric and spectroscopic hardware, as well as
the data reduction procedures and targeting strategy, can be found in Gunn et
al. (1998, 2006), Lupton et al. (1999, 2001), Pier et al. (2003), and
Stoughton et al. (2002).

Followup observations of the new magnetic candidates utilized circular
polarimetry with the instrument SPOL (Schmidt et al. 1992) attached to the
Steward Observatory 2.3 m Bok and 1.5 m Kuiper telescopes on Kitt Peak and Mt.
Bigelow, respectively. In the spectroscopic configuration used, the
1200$\times$800 SITe CCD provides a coverage $\sim$$\lambda\lambda4000-8000$
and resolution $\sim$15~\AA.  Data were obtained in a series of 16 min
observational sequences, each sequence yielding circular polarization and total
spectral flux as functions of wavelength.  The results confirm the cyclotron
nature of the humps, as shown in Figures 2 and 3.  Imaging polarimetry was
acquired of SDSS~J1031+2028 with the same instrument, using a plane mirror
instead of the grating and replacing the entrance slit by a square aperture
that admits a 1 arcmin square region of the sky on the 1.5 m and 2.3 m
telescopes.  Further details of the instrumentation and analysis procedures can
be found in Schmidt et al. (1992) and S05.

Finally, a run of CCD photometry on SDSS~1031+2028 was obtained with the USNO
1.3~m telescope in unfiltered light within a few weeks of its discovery.  The
object was a difficult target for this small telescope and the end of the 7.5
hr run was plagued by clouds, but a periodic variation is apparent in the
data.  The observational histories for the two new objects are summarized in
Table 1.

\section{New Low Accretion-Rate Magnetic Binaries}

A specific accretion rate $\dot m \lesssim 10^{-2}$ g cm$^{-2}$ s$^{-1}$ onto a
strongly magnetic white dwarf characterizes the ``bombardment'' regime where
incident ions lose their kinetic energy via small-angle scatterings in the
atmosphere rather than through a hydrodynamic shock.  The electrons radiate
efficiently in the magnetic field, and the resulting low plasma temperature
produces cyclotron emission that is confined to only a few low well-defined
harmonics (Woelk \& Beuermann 1996).  With total accretion rates $\dot M
\lesssim 10^{-13}~M_\sun$ yr$^{-1}$, these magnetic systems are virtually
nonexistent in X-rays and can only be discovered through deep optical surveys.

\subsection{SDSS~J1059+2727}

The spectrum of SDSS~J1059+2727 displays obvious humps centered around
$\lambda$6400 and $\lambda$4750 that can be identified with cyclotron harmonics
$m=3$ and 4, respectively in a field of 57 MG.  The predicted location of the
$m=2$ harmonic is then just off the spectrum at $\lambda \approx 9500$~\AA.
With broadband colors of $u-g=+1.27$, $g-r=+1.84$ (Table 2), the object falls
squarely on the simulation track of low-$\dot M$ systems in the color-color
plane at the correct field strength (Figure 1 of S05).  The {\it psf\/}
magnitude of $g=22.1$ is 0.5 mag brighter than the fiber magnitude obtained
more than a year later, but orbital brightness modulations are expected, and
thus far the object has always been a challenging spectroscopic target on a
modest telescope. Our nearly 2 hr of spectropolarimetric observations on the
Bok 2.3~m reflector measured substantial overall circular polarization in each
polarimetric sequence, $v=-8$ to $-$12\%, but there appears to be no coherent
dependence on time that would indicate a spin/orbital period\footnote{The
circularly polarized flux in the (strongest) $m=3$ harmonic is equivalent to
an $R$ magnitude of 24.1.}. Furthermore, even though \halpha\ can be
recognized in emission in several observations, it is so weak that we can only
set an upper limit of $\pm$100 km s$^{-1}$ on any Doppler variation through
the series.  We infer that the binary probably has a relatively long orbital
period, $P>3$ hr, and/or has a low orbital inclination.

Despite 98 min of total integration with the SDSS spectrograph, it is
difficult to assign a spectral type to the secondary in SDSS~J1059+2727 (see
Figure 1), given the combination of a relatively low signal-to-noise ratio and
significant contamination produced by the two cyclotron humps.  A comparison
with SDSS main-sequence spectral standard stars (Hawley et al. 2002) suggests a
best match to an M4 spectral type, with an uncertainty of about 1 subclass.
With this result, the measured flux in the interval around $\lambda=7500$~\AA,
which is between cyclotron harmonics, can be used together with the
spectrophotometric calibration of M-stars from the appendix of S05 to estimate
a distance to SDSS~J1059+2727.  The result ranges from $D\sim460$ pc for an M5
secondary to 1300 pc for an M3 star.  A distance estimate allows the accretion
luminosity to be computed, assuming that cyclotron emission dominates the
light output.  The narrowness of the harmonics,
$\Delta\lambda/\lambda\sim0.05$, demonstrates that the plasma temperature is
low, $kT_e\sim1$ keV, leading to the prediction that X-ray emission from
SDSS~J1059+2727 will be negligible.  If we use the model calculations of
Ferrario et al. (2002, unpublished poster at the Cape Town IAU Colloquium
\#190 on magnetic cataclysmic variables; see also Ferrario et al. 2005) as a
guide to the amount of radiation at unseen harmonics, we obtain an estimate
for the accretion luminosity of $L_{\rm acc} = 0.4-3\times10^{30}$ erg
s$^{-1}$, or a total mass transfer rate of $\dot M=0.6-5\times10^{-13}~M_\sun$
yr$^{-1}$. This is nearly 3 orders of magnitude below the accretion rates of
Polars during high states and comfortably among the rates determined for the
previous low-$\dot M$ examples (see, e.g., S05).

Applying the same method as was used for SDSS~J1553+5516 by Szkody et al.
(2003), the observed flux between the bluest harmonics can be used to place an
upper limit on the temperature of the white dwarf.  For SDSS~J1059+2727 we
allow a rather generous level of $F_\lambda=2\times10^{-18}$ erg cm$^{-2}$
s$^{-1}$ \AA$^{-1}$ in the range $4300-4500$~\AA\ (Figure 1), and compare this
to fluxes from Bergeron et al.'s (1995) nonmagnetic $\log~g = 8$ DA models for
the typical white dwarf mass of 0.6~$M_\sun$ ($R_{\rm wd}=0.012~R_\sun$). The
result is $T_{\rm wd}\le5500$~K for $D=460$~pc and $\le$8500~K for the distant
limit. These and other parameters derived for SDSS~J1059+2727 are entered in
Table 2.

\subsection{SDSS~J1031+2028}

Three prominent humps near $\lambda\lambda$8750, 6400, 5300 in the spectrum of
SDSS~J1031+2028 are readily identified with harmonics 3$-$5 in a field of 42
MG. The two higher harmonics are confirmed by the spectropolarimetric
observations to be strongly circularly polarized and thus cyclotron in nature.
Indeed, the only significant difference between the spectra in Figures 1 and 3,
which are separated by 3 months, is the disappearance of what was already a
very weak \halpha\ emission line in the SDSS data (upon inspection, the
apparent line at 7561~\AA\ in the survey spectrum is found to be an
uncorrected cosmic ray).  The strengths and shapes of the cyclotron features
clearly vary periodically over the spectropolarimetric sequence.  This is
reflected in the time dependence of the spectrum-summed degree of circular
polarization shown as Figure 4.  Here, the fitted sine curve has a period
$P=1.37 \pm 0.10$ hr = 0.057 $\pm$ 0.004 d, which we take to be the spin
period of the white dwarf.  The varying cyclotron component also appears as a
brightness modulation, as is evident in the unfiltered CCD differential
photometry obtained at the USNO 1.3~m telescope and shown in Figure 5. The
least-squares sine fit to the light curve (shown) has a semiamplitude of 0.28
mag. Although the data are not of the highest quality due to variable cloud
cover during the observations, the run covers several cycles and improves upon
the polarimetric period with $P=0.0578\pm0.0016$~d. Of course, without radial
velocity information, we cannot be sure that this period is also the orbital
period, but all previous low-$\dot M$ systems have been synchronized (S05) and
the value certainly falls in the appropriate range. At all observing epochs
spanning more than 1 yr, the object has been fainter than $g=18$.

New among the eight low-$\dot M$ systems discovered thus far is the lack of
evidence for a late-type star in the optical spectrum of SDSS~J1031+2028.
Instead, a blue continuum underlies the cyclotron features. The
spectropolarimetric sequence reveals no periodic variation in the flux between
harmonics (here, measured at 5000~\AA) to a limit of $\pm$15\%, so we take the
continuum to be from the integrated disk of the white dwarf, as opposed to a
heated accretion spot\footnote{A hot spot that covers only a small portion of
the stellar disk would imply a smaller distance and lead to a much lower
implied mass transfer rate than we estimate below.}.  For a dipolar field
strength of 42 MG, the expected surface-averaged magnetic field on the white
dwarf is $\sim$30 MG, depending on limb-darkening and inclination.  The
predicted locations of the principal Zeeman absorption features of \halpha\ and
\hbeta\ are shown for this field strength below the observed spectrum in Figure
1, and the correspondence between several of the \hbeta\ components with narrow
dips shortward of 5000~\AA\ supports our interpretation of the underlying
continuum as being that of the white dwarf.  \halpha\ is more sensitive than
\hbeta\ to the magnetic field strength and at a temperature of $\sim$9500~K
(derived below), considerably weaker than \hbeta\ in normal DA stars, so it is
not surprising that a similar correspondence is absent for \halpha.

Again we compare the slope of the underlying component with $\log~g = 8$ DA
models of Bergeron et al. (1995) to estimate the white dwarf temperature in
SDSS~J1031+2028.  Shown as crosses and plus signs in the lower panel of Figure
1 are computed flux distributions at 100~\AA\ intervals for nonmagnetic white
dwarf continua with $T_{\rm eff} = 8000$ K and 11000 K, respectively,
normalized to the observed flux in the gap between cyclotron harmonics at
6000~\AA.  The 8000 K model falls short of the observed flux at nearly all
wavelengths short of 5000~\AA, while the 11000 K model significantly
underpredicts the light between $m=3$ and 4 around 7500~\AA.  We take these
models as indicative of the range in temperature for the underlying white
dwarf and quote $T_{\rm wd}=9500\pm1500$ K.  The normalization of the spectral
flux then implies $D=270-430$ pc. If we again take cyclotron emission to be the
dominant energy loss mechanism, we find that $L_{\rm acc}=1-3\times10^{30}$
erg s$^{-1}$, and $\dot M=1.5-4\times10^{-13}~M_\sun$ yr$^{-1}$.  Of course,
satellite observations should be carried out to verify the assumption that
cyclotron cooling dominates by ruling out significant X-ray emission from both
of these new magnetic systems. We note here that neither object corresponds to
a {\it ROSAT\/} source.

The measured flux between the reddest harmonics can now be used to constrain
the nature of the secondary star in SDSS~J1031+2028.  In the $7400-7900$~\AA\
gap, the observed flux is $1.4\times10^{-17}$ erg cm$^{-2}$ s$^{-1}$
\AA$^{-1}$.  We again use the absolute spectrophotometry of main-sequence stars
with accurate parallaxes from S05, and find that the absence of a detectable
molecular bandhead at 7600~\AA\ rules out a main-sequence secondary earlier
than M7 for the 270 pc distance, or earlier than M6 if the system is at the
maximum 430 pc.  It is useful to note that a main-sequence secondary earlier
than M6 is also precluded by the fact that it cannot overfill its Roche lobe in
the $P=1.37$ hr binary. These and other characteristics for SDSS~J1031+2028
are listed in Table 2.

\section{Implications and Conclusions}

With the discovery of each new low-$\dot M$ binary that cannot be identified
with a prior X-ray detection or variable star, it is becoming increasingly
clear that the accretion rates measured for many of these systems are relevant
over long time intervals and thus are more indicative of stellar wind mass
loss than Roche-lobe overflow.  All eight binaries to date yield rates in the
range $5\times10^{-14}-3\times10^{-13}~M_\sun$ yr$^{-1}$, with no apparent
dependence on orbital period or spectral type of the donor star. These values
are $\sim$$2-10$ times the current solar wind mass loss rate of
$2\times10^{-14}~M_\sun$~yr$^{-1}$ (Hundhausen  1997), but solar variability on
periods from 11 to 2300 yr is inferred (Sonett et al. 1997).  Since previous
attempts to measure the mass loss rates from low-mass main-sequence stars have
provided only upper limits in the range $\sim$$10^{-13} - 10^{-11}~M_\sun$
yr$^{-1}$ (e.g., Lim \& White 1996; Wargelin \& Drake 2001), the accretion
rates of the low-$\dot M$ magnetic binaries may actually prove to be the first
realistic measurements of stellar mass loss at the cool end of the main
sequence.  Of course, the relevance of these numbers presumes that the magnetic
siphon process is highly efficient (Li et al. 1995), and that wind mass loss is
not strongly affected by the proximity of the white dwarf or forced rotation
of the secondary.

A peculiar trait of the first six low-$\dot M$ magnetic systems (S05) was their
sharing of a nearly common magnetic field strength, $B=60-68$ MG.  Model flux
distributions, computed as a function of magnetic field strength by S05, were
projected onto the $u-g$, $g-r$ plane to show that, even though fields near 60
MG were prone to targeting of an SDSS spectroscopic fiber, other ranges in $B$
were also susceptible, and it was suggested that discoveries at both
significantly higher and lower field strengths should appear.  While the 57~MG
field on SDSS~J1059+2727 is near the above range, SDSS~J1031+2028 at 42 MG
verifies that prediction. Ironically, however, the broadband photometry of
SDSS~J1031+2028 is dominated by the contribution of the white dwarf (as well
as the lack of a cool secondary component), so its colors fall nearer the
white dwarfs in the color-color plane than the simulation track for the
measured field strength (Figure 1 of S05).  In any case, it appears that a
$\sim$40 MG magnetic field on a white dwarf in a very short period binary is
sufficient to efficiently engineer the magnetic siphon that channels the
stellar wind onto the poles.

Earlier work has characterized the low-$\dot M$ magnetic systems as pre-Polars
because of the very low measured accretion rates, secondary stars that
underfill their Roche lobes, and relatively cool white dwarfs (S05; Schmidt
2005; Webbink \& Wickramasinghe 2005).  While these arguments are compelling in
general, there exists the possibility that one or more of the eight members
cataloged to date are nearer the opposite evolutionary extreme.  That is, they
might be temporarily interrupted or even extinct Polars.  A well-known example
of the former is EF Eridani, which was a bonafide Polar until falling into a
state of near-zero accretion in 1997 (Harrison et al. 2004 and references
therein), and it has only been within the past year that EF Eri has been seen
in an active state (Howell et al. 2006).  In fact, with an orbital period of
81 min, a primary at $T_{\rm wd}=9500$~K; Beuermann et al. 2000), and a
brown-dwarf secondary star, EF Eri bears a striking resemblance to
SDSS~J1031+2028.  If an interrupted Polar explanation is to be applied to a
nearby low-$\dot M$ binary like SDSS~J1553+5516 ($D\sim130$~pc; S05), the
current low state must have been in place for decades in order that the system
escape discovery as a variable star (e.g., AM Her, VV Pup) or a strong X-ray
source in the all-sky surveys.  The low-state duration would outstrip even that
displayed by EF Eri.  Because of accretion-induced heating (e.g., Townsley \&
Bildsten 2004), the interrupted Polar explanation should be less viable for the
low-$\dot M$ magnetic binaries that contain cool white dwarfs.  Examples like
SDSS~J1324+0320 and SDSS~J2048+0050 ($T_{\rm wd} \lesssim 7500$K; S05) and
SDSS~J1059+2727 ($T_{\rm wd} < 8500$K) are cooler than the coolest white dwarfs
in Polars (Sion 1999; Araujo-Betancor et al. 2005).  However, if we take the
cooling curve of WZ Sagittae as a guide (Godon et al. 2006), any object that
has been in a continuous low state for $\gtrsim$3 yr can be considered to
contain a white dwarf very near the evolutionary temperature for its age.  This
is especially true for magnetic accretion binaries, which tend to have
below-average accretion rates for their orbital periods and spend a significant
fraction of the time in states of weak accretion (Ramsay et al. 2004).

Ambiguity over the state of evolution need not impact claims of wind accretion,
however, since $H$ and $K$-band brightness modulations during the protracted
low-state of the interrupted Polar EF Eri also appear to require low-harmonic
cyclotron emission from poles accreting in the bombardment regime (Harrison et
al.  2004).  In addition, the detached $\sim$90~min period white dwarf + brown
dwarf binary SDSS~J121209.31+013627.7, whose primary has $T_{\rm
wd}=10,000\pm1000$~K and a dipolar magnetic field of $B_p\sim13$ MG (Schmidt et
al. 2005b), can be regarded as a system closely related to the low-$\dot M$
magnetic binaries, now that $K$-band photometry (Debes et al. 2006) has found
evidence for cyclotron emission fed by a wind from the L7 companion.

SDSS~J1212+0136 was originally identified as a binary from the presence of weak
Balmer emission lines produced by radiative heating and/or activity at the
secondary's surface.  Only white dwarf + brown dwarf systems with the
combination of a reasonably hot primary star and short orbital period are
likely to show this effect, and the difficulty in identifying white dwarf +
brown dwarf pairs by any technique is well-documented (e.g., Debes et al.
2004; Farihi et al.  2005).  The discovery of the L7 component in
SDSS~J1212+0136 follows only GD 165 (DA + L3$-$4; Becklin \& Zuckerman 1988)
and GD 1400 (DA + L6:; Farihi \& Christopher 2004).  While the relatively
modest 13~MG magnetic field on SDSS~J1212+0136 constrains the emitting low
cyclotron harmonics to the IR, optical cyclotron emission induced by accretion
of the stellar wind by a strongly magnetic white dwarf offers another, possibly
sensitive avenue for identifying binaries with low-mass companions and periods
up to several hours.  With a thus-far undetected secondary, SDSS~J1031+2028 may
prove to be the such first example.  Allowance for the efficiencies of surveys
in discovering low-$\dot M$ magnetic systems and for the fraction of white
dwarfs that are magnetic (e.g., Liebert et al. 2003; Kawka et al. 2003) should
permit the statistics of magnetic samples to be extended to white dwarf + brown
dwarf pairs without regard for magnetism.

Spectral synthesis and targeting simulations by S05 revealed that even
sophisticated multicolor surveys like the SDSS are blind to several field
strength regimes.  When these selection effects were crudely taken into
account, it was found that, even though the current sample is small in
number, the space density of pre-Polars may be quite significant.  Indeed,
wind-accretion magnetic binaries are revealing mass transfer rates so low that,
not only are they virtually invisible in X-rays, but the cyclotron emission
component is little brighter than either of the stellar continua.   Binaries
with still lower accretion rates will eventually be lost in the stellar locus
regardless of field strength and fail the spectroscopic targeting algorithms.
When an accurate census is finally taken, wind-accreting magnetic binaries may
actually prove to be more abundant than the classical X-ray emitting,
Roche-lobe overflow Polars.

\acknowledgements{The authors are grateful to P. Smith for assistance at the
telescope and J. Bochanski for helping to type the spectra of cool main
sequence stars.  AAH thanks the USNO-Flagstaff Station Director for the time
allocation on the USNO-FS 1.3~m telescope. Funding for the SDSS and SDSS-II has
been provided by the Alfred P. Sloan Foundation, the Participating
Institutions, the National Science Foundation, the U.S. Department of Energy,
the National Aeronautics and Space Administration, the Japanese Monbukagakusho,
the Max Planck Society, and the Higher Education Funding Council for England.
The SDSS Web Site is http://www.sdss.org/. The SDSS is managed by the
Astrophysical Research Consortium for the Participating Institutions. The
Participating Institutions are the American Museum of Natural History,
Astrophysical Institute Potsdam, University of Basel, Cambridge University,
Case Western Reserve University, University of Chicago, Drexel University,
Fermilab, the Institute for Advanced Study, the Japan Participation Group,
Johns Hopkins University, the Joint Institute for Nuclear Astrophysics, the
Kavli Institute for Particle Astrophysics and Cosmology, the Korean Scientist
Group, the Chinese Academy of Sciences (LAMOST), Los Alamos National
Laboratory, the Max-Planck-Institute for Astronomy (MPIA), the
Max-Planck-Institute for Astrophysics (MPA), New Mexico State University, Ohio
State University, University of Pittsburgh, University of Portsmouth, Princeton
University, the United States Naval Observatory, and the University of
Washington. Support is provided by the NSF for the study of magnetic stars and
stellar systems at the University of Arizona through grant AST 03-06080, and
for cataclysmic variables at the University of Washington through AST
02-05875.}

\clearpage

\begin{deluxetable}{lcclcc}

\tablecaption{LOG OF OBSERVATIONS}

\tablewidth{5.5truein}

\tablehead{\colhead{Object} &
\colhead{UT Date} &
\colhead{Telescope} &
\colhead{Type} &
\colhead{Duration} &
\colhead{Cir. Pol.} \\
\colhead{(SDSS+)} &
\colhead{(yyyymmdd)} &
\colhead{} &
\colhead{} &
\colhead{(h:mm)} &
\colhead{(\%)} }

%% All data must appear between the \startdata and \enddata commands
\startdata
J1031+2028 & 20050310 & SDSS 2.5~m & Imaging     & $\cdots$ & $\cdots$ \\
                    & 20060204 & SDSS 2.5~m & Spectros.   & 0:48 & $\cdots$ \\
                    & 20060226 & USNO 1.3~m & Imaging     & 7:28 & $\cdots$ \\
                    & 20060502 & Bok 2.3~m  & Spectropol. & 2:37 & ~~$-$4 to $-$22
\\
J1059+2727 & 20041215 & SDSS 2.5~m & Imaging     & $\cdots$ & $\cdots$ \\
                    & 20060306 & SDSS 2.5~m & Spectros.   & 1:38 & $\cdots$ \\
                    & 20060402 & Kuiper 1.5~m & Imaging Pol.  & 1:13 & $-$7.9 \\
                    & 20060503 & Bok 2.3~m  & Spectropol. & 1:52 & $-$3.1
\enddata

%% Include any \tablenotetext{key}{text}, \tablerefs{ref list},
%% or \tablecomments{text} between the \enddata and
%% \end{deluxetable} commands

%% No \tablecomments indicated

%% No \tablerefs indicated

\end{deluxetable}

\begin{deluxetable}{lcc}

\tablecaption{PROPERTIES OF NEW MAGNETIC BINARIES}

\tablewidth{4.05truein}
\tablehead{\colhead{} &
\colhead{SDSS J1031+2028} &
\colhead{SDSS J1059+2727}
}

%% All data must appear between the \startdata and \enddata commands
\startdata
Plate-MJD-Fiber & 2375-53770-636 & 2359-53800-051 \\
$g$ & ~18.26 & ~22.09 \\
$u-g$ & +0.09 & +1.27 \\
$g-r$ & $-$0.27 & +1.84 \\
$r-i$ & $-$0.36 & +0.35 \\
$i-z$ & $-$0.15 & +1.05 \\
$P$ (h) & 1.37 &  $>$3: \\
$B$ (MG) & 42  & 57 \\
$\dot M$ ($\times 10^{-13} M_\sun$ yr$^{-1}$) & $1.5-4$ & $0.6-5$ \\
$D$ (pc) & $270-430$ & $460-1300$ \\
$T_{\rm wd}$ (K) & $9500\pm1500$ & $\le$8500 \\
Sp. Type (secondary) & $\ge$M6 & M3 $-$ M5 \\
\enddata

%% Include any \tablenotetext{key}{text}, \tablerefs{ref list},
%% or \tablecomments{text} between the \enddata and
%% \end{deluxetable} commands

\end{deluxetable}

\clearpage

\begin{figure}
\includegraphics{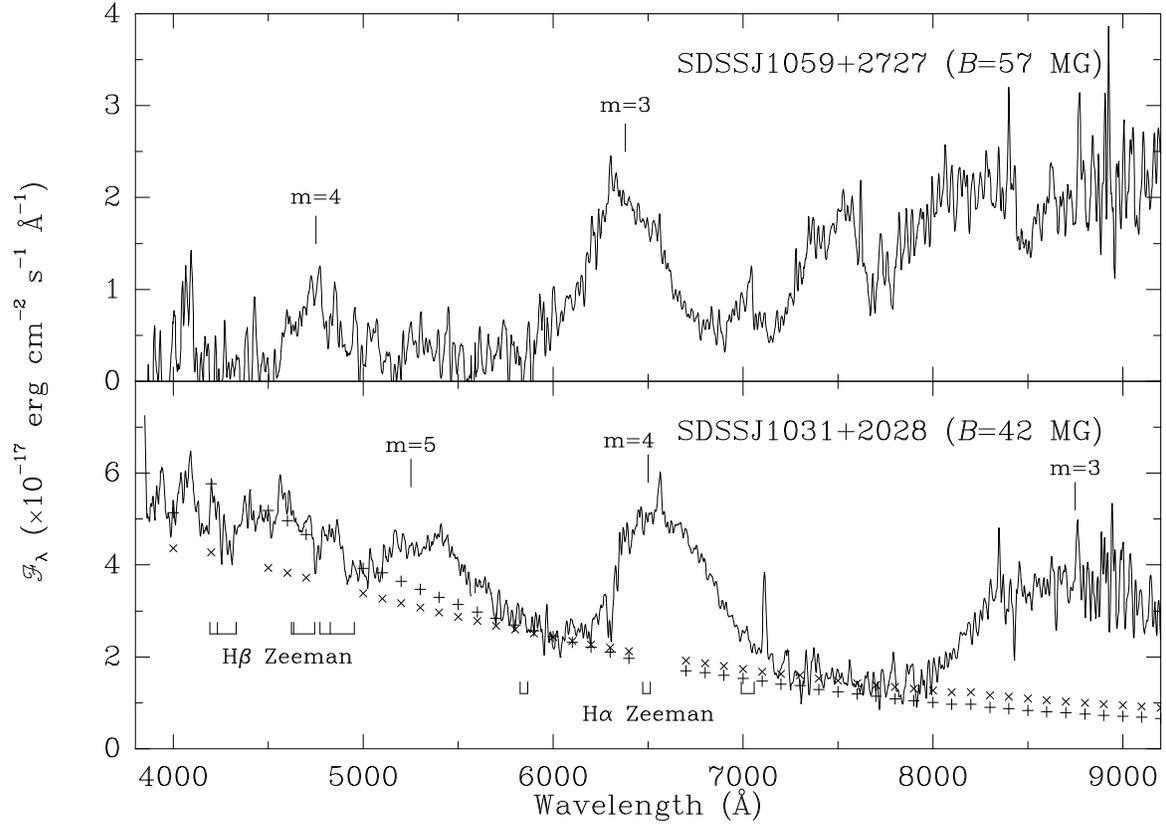}
\vspace{3.truein}

\figcaption{SDSS spectra of the two new low-$\dot M$ magnetic binaries, shown
at a resolution of $\sim$8~\AA. Cyclotron harmonics are indicated. The $m=2$
harmonic in SDSS~J1059+2727 is expected to be centered near 9500~\AA.  Shown
below the spectrum of SDSS~J1031+2028 are the predicted locations of \hbeta\
and \halpha\ photospheric Zeeman components for an estimated surface-averaged
field strength of 30 MG.  Also shown are normalized model atmosphere flux
distributions for $\log~g = 8$ nonmagnetic DA white dwarfs with $T_{\rm
eff}=8000$ K {\em (crosses)\/} and $T_{\rm eff}=11000$ K {\em (pluses)}. These
indicate the estimated range in surface temperature for the underlying white
dwarf.}
\end{figure}

\clearpage

\begin{figure}
\includegraphics{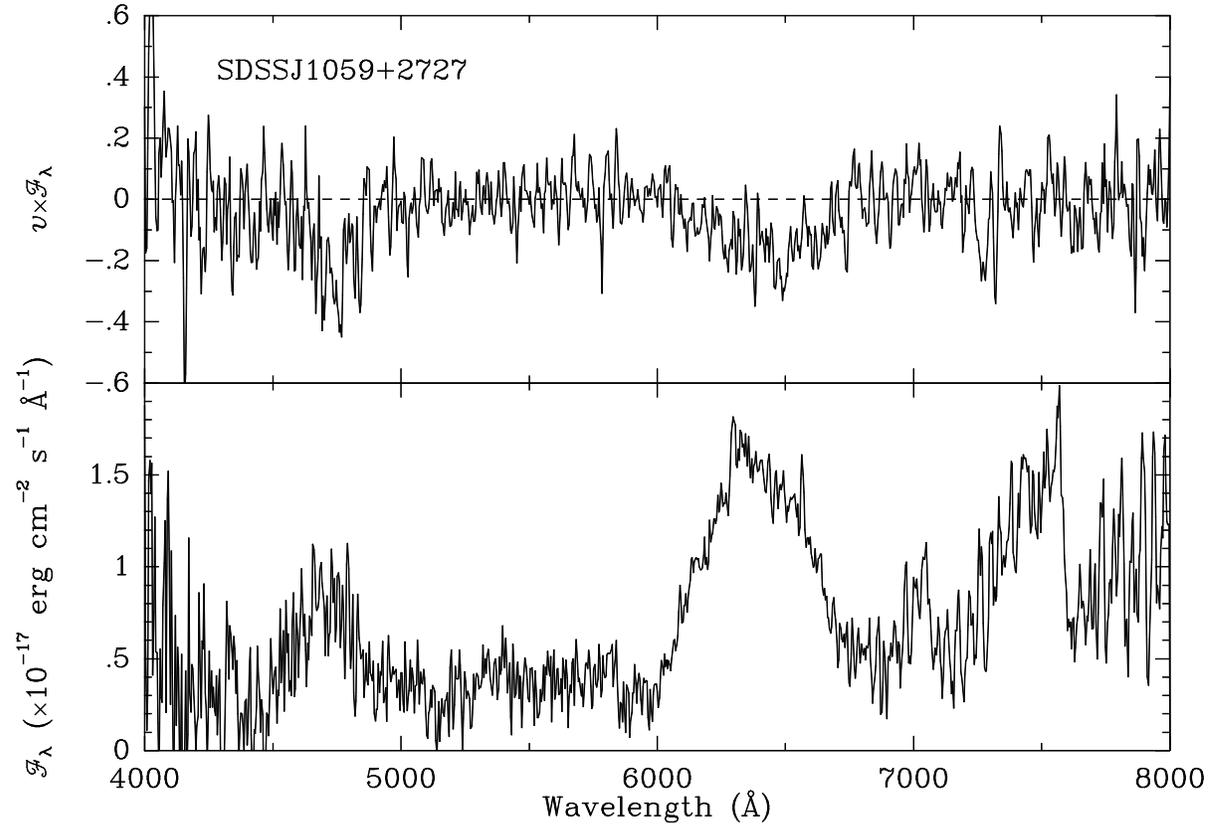}
\vspace{3.truein}

\figcaption{Circularly polarized flux ($v\times F_\lambda$) and total flux
spectra for SDSS~J1059+2727 obtained with a resolution of $\sim$15~\AA\ at the
Bok 2.3~m telescope in 2006 May.}
\end{figure}

\clearpage

\begin{figure}
\includegraphics{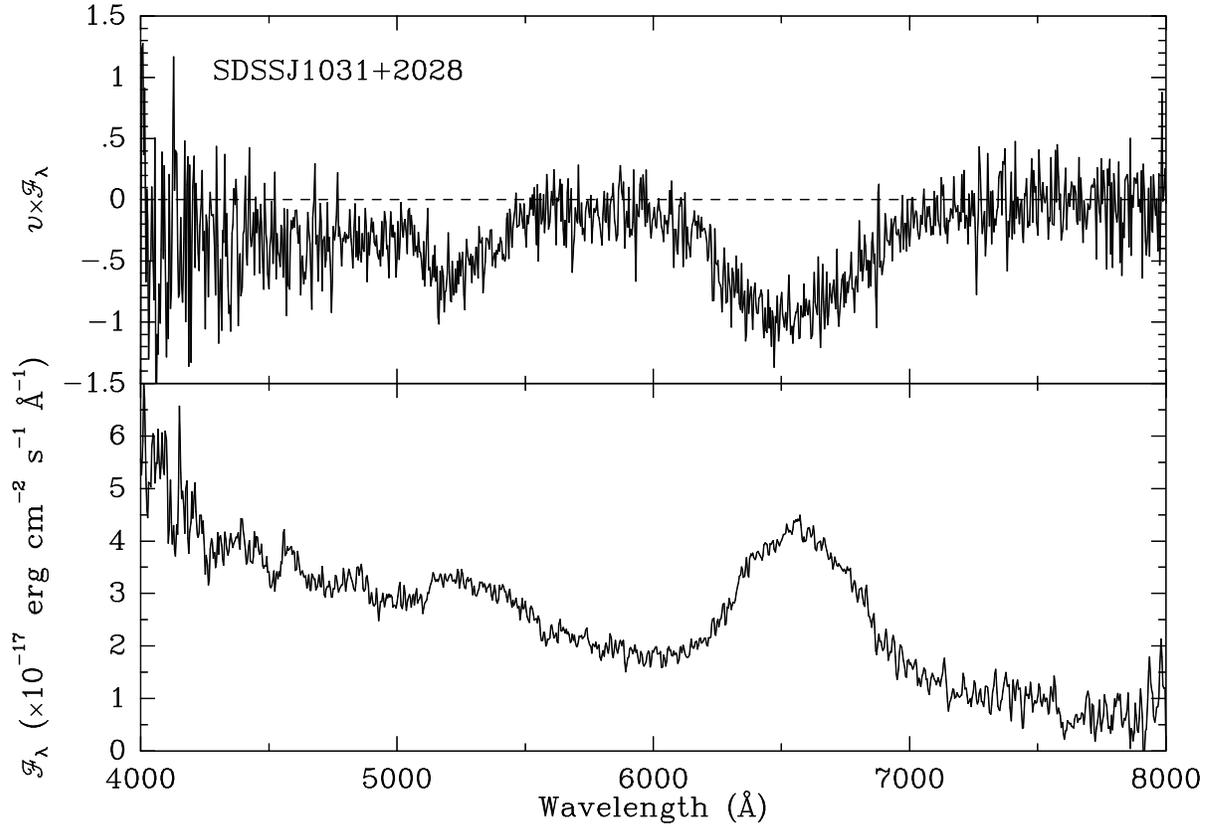}
\vspace{3.truein}

\figcaption{As in Figure 2 for SDSS~J1031+2028.}
\end{figure}

\clearpage

\begin{figure}
\includegraphics{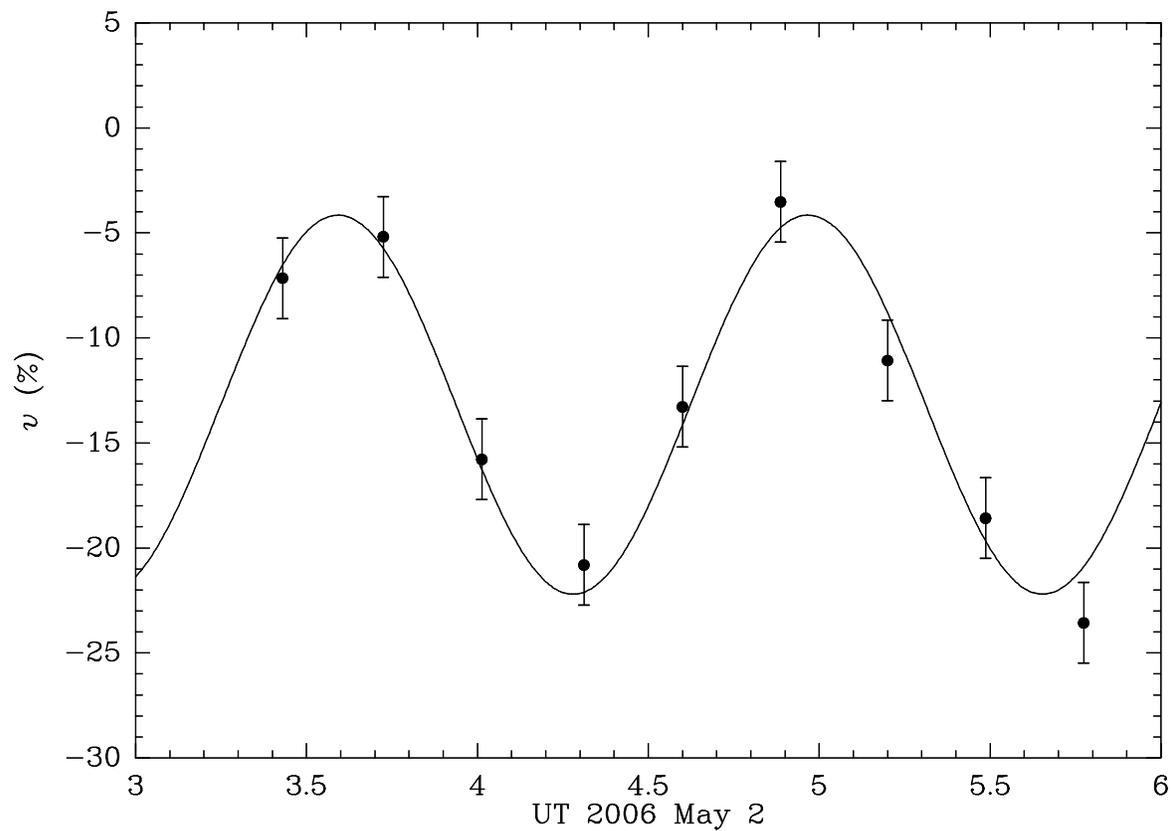}
\vspace{3.truein}

\figcaption{Time dependence of circular polarization for SDSS~J1031+2028,
coadded over the optical spectrum.  The best-fit sinusoid (shown) has a period
of 0.057$\pm$0.004 d.}
\end{figure}

\clearpage

\begin{figure}
\includegraphics{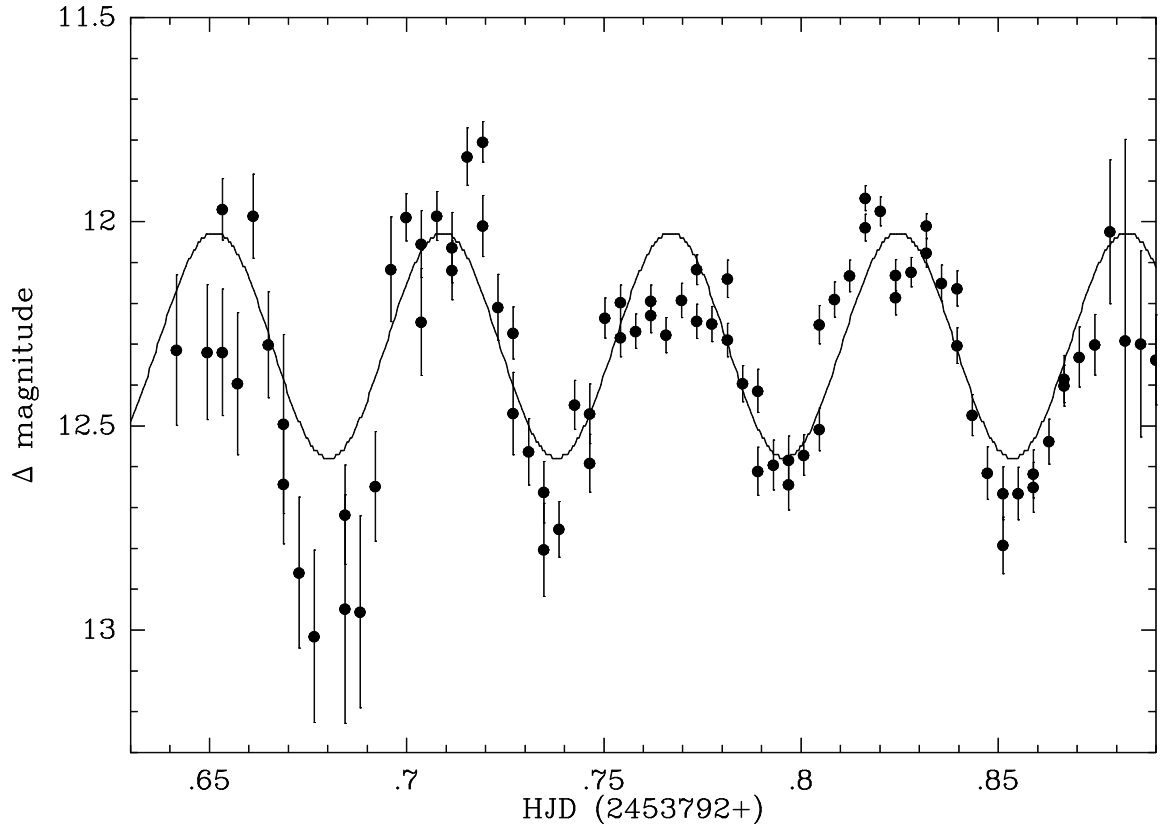}
\vspace{3.truein}

\figcaption{Light curve of SDSS~J1031+2028 obtained with an unfiltered CCD on
the USNO 1.3 m telescope. Magnitudes are relative to  comparison stars in the
same data frames.  The fitted sinusoid has a semiamplitude of 0.28 mag and
period of $0.0578\pm0.0016$ d.}
\end{figure}
\end{document}